# X–ray clusters: towards a new determination of the density parameter of the universe

J. Oukbir[1] and A. Blanchard[2]

[1]CE–Saclay DSM/DAPNIA/SAp, Orme des Merisiers, F–91191 Gif–sur–Yvette Cedex

[2]Observatoire Astronomique de Strasbourg, 11 rue de l'Université, F–67000 Strasbourg

November 12, 1996

**Abstract.** We use a self-consistent modeling of X–ray cluster properties to constrain cosmological scenarios of structure formation in the case of open cosmological models. We first show that an unbiased open model can reproduce present day observations, provided that the density parameter is in the range $0.15 - 0.4$. Although this estimate is derived in a rather different way, it is very close to dynamical measurements. We also obtain constraints on the shape of the initial fluctuation power spectrum: we find that the local spectral index $n$ falls in the range $-1.1 \leq n \leq -1.6$ on scales between 10 and $30 h^{-1}$Mpc. We focus on differences between open models and the Einstein–de Sitter model, and we exhibit a test which depends only on the cosmological parameter and is insensitive to the shape of the initial fluctuations power spectrum: we show that, contrary to what is generally admitted, the redshift distribution of X–ray clusters selected on the basis of their apparent temperature is almost independent of the power spectrum and that it is only sensitive to the value of $\Omega_0$. Since the relation between the virial mass and the X–ray temperature seems rather robust, we expect very little theoretical uncertainty in the modeling. This allows us to show that the combined knowledge of the redshift distribution of X–ray flux selected clusters and that of the luminosity–temperature correlation at high redshifts offers a new and efficient way to measure the mean density of the universe. This appears to be a clean test of $\Omega_0$, not suffering of the problems attached to local density estimators.

We have tentatively applied our test to the redshift distribution of clusters detected within the Einstein medium sensitivity X–ray survey. As temperature information is not available for these clusters, no definitive conclusion can be reached. Taking into account the selection function, we find that the observed redshift distribution can be equally well fitted with a non–evolving $L_X - T$ correlation in the case of the critical universe or a substantial negative evolution in the case of a low–density universe. Our test appears then to be able to provide a definitive answer to the question concerning the density of the universe, provided that the $L_X - T$ relation is measured at a redshift of the order of $0.4 - 0.5$ and that temperature measurements can be achieved with an accuracy of the order of 20%.

# 1. Introduction

The determination of the density parameter, $\Omega_0$, has received much attention in cosmology. Theory favors a value of $\Omega_0$ equal to one, because inflation naturally leads to this value. However, observations indicate a much lower value of $\rho_0$. The contribution of the visible part of galaxies is $\Omega_0 \sim 0.004$ (Peebles, 1993). If massive galactic halos are taken into account, then the derived value of $\Omega_0$ is about 3 to 10 times higher. On larger scales, from dynamical estimates within galaxy clusters as well as from the cosmic virial theorem, the measured value of $\Omega_0$ is around 0.2 (for a review on estimates of the content of the universe, see White 1990). However, it must be emphasized that the latter dynamical derivations of the mean density are based on the assumption that the galaxy systems considered are representative of the universe. This is probably not the case: different scales imply different values of $\Omega_0$, with a tendency for the inferred mean density to increase with scale. On scales of a few tens of Mpc, the value derived from the peculiar velocity field is consistent with the critical value (Dekel 1993). The reason for this discrepancy is not yet fully understood. It is therefore of fundamental interest to estimate the density of the universe by methods which differ from the usual dynamical ones. The goal of our paper is to present such a possibility.

There are recent attempts to constrain the density parameter from the morphology of galaxy clusters. Assuming that the distribution of density perturbations at high redshift is gaussian, Richstone *et al.* (1992) placed a lower limit on $\Omega_0$ from an estimate of the fraction of rich clusters with significant substructures. Their study is based on the expectation that most structures have settled into some stable state in a low density universe. However, such an argument relies on subtle dynamics in the non-linear regime and should be interpreted with caution. Although the conclusions are still preliminary, a similar argument based on the structure of the X-ray gas have also led to a high value of the density parameter of the universe (Mohr *et al.*, 1995).

As we have mentioned, although most of the observations do not provide support for such a conclusion, many theorists have favored a value of $\Omega_0$ equal to 1, primarily because of the attractive features of inflationary models. Furthermore, inflation also predicts the shape of the power spectrum of the primordial density fluctuations generated during the very early universe and provided that the nature of the dark matter is specified, the amplitude of the power spectrum remains the only free parameter in these models (at least, in their simpler version). In the case of a low density open model, and in the absence of a standard scenario, the fluctuations are not well specified and the more conservative approach is to assume on a given scale range, a power spectrum of the form $P(k) \propto k^n$, where $n$ is a free parameter. The index $n$ as well as the normalization of the power spectrum are then to be constrained. The abundance of galaxy clusters provides a interesting tool to derive these parameters on scales of a few tens of Mpc, since clusters result from the collapse of fluctuations with typical comoving radius in this range.

For practical purposes, the Press & Schechter formalism (1974) (PS hereafter) gives an accurate determination of the mass function. Provided that the mass is related to some observed quantity like the X-ray temperature or the X-ray luminosity, a comparison with observations allows one to constrain the amplitude and the shape of the power spectrum on galaxy cluster scales. In the recent years, many authors have investigated for the case of a critical universe, the consequences of the observed X-ray temperature and X-ray luminosity distribution functions. Oukbir *et al.* (1996) (OBB hereafter) provide a comprehensive discussion of the subject and evaluate in detail the various constraints that can be inferred from such a study. The case of a low density universe has also been investigated, but restricted to the context of a flat universe or with the shape of the power spectrum given by the CDM theory (Efstathiou *et al.* 1992, Lilje 1992, White *et al.* 1993, Bartlett & Silk 1993, Liddle *et al.* 1995), or to address some specific topic (Oukbir & Blanchard 1992, Hattori & Matsuzawa 1995, Eke it et al. 1996).

In this paper we extend these studies to the case of a low density open universe. In Section 2, we discuss the arguments supporting the PS formalism and the relations between the cluster mass, the X-ray temperature and the X-ray luminosity in low density universes. In Section 3, we constrain the shape and the amplitude of the initial mass power spectrum by using the observed temperature distribution function. Since the normalization of the power spectrum depends on $\Omega_0$, we also constrain the density parameter. The critical model and the open model both reproduce the observations at $z = 0$ and cannot therefore be distinguished by mean of present day data. However, in open low density universes structure formation is occuring much earlier than in flat models. Therefore, we also examine the observational constraints implied by high redshift clusters. In Section 4, we investigate the redshift distribution of X-ray temperature selected galaxy clusters. We show that this distribution provides a robust information on the mean density of the universe independently of the power spectrum of the fluctuations. During the preparation of this work, the redshift distribution of the Einstein clusters has been established. We have then examined the possibility of constraining the density parameter from these data, despite the fact that we do not have information on the temperature of these clusters.

## 2. The theoretical temperature distribution function

*2.1. The mass function*

The formation of nonlinear bound objects is expected to occur from the collapse of initial density fluctuations that grow under the influence of gravity. In the following, these fluctuations are assumed to obey gaussian statistics, although in the low-$\Omega_0$ case the motivation for non-gaussian fluctuations is higher. The case of non-gaussian fluctuations has been investigated in OBB and the difference with the gaussian case was found rather weak in front of changes introduced by changing $\Omega_0$. Non-linear structures of mass $M$ are expected to form from initial density fluctuations of comoving size $R_s$ that contain mass $M$. We parameterize the initial mass power spectrum on galaxy clusters scale by a power law:

$$P(k) \propto k^n.$$

The variance of the initial density contrast field $\delta(\mathbf{x})$, smoothed with some window function at scale $R_s$, is then:

$$\sigma_i(M) \propto M^{-(n+3)/6}.$$

In the case of a top-hat function, the window mass is

$$M = \frac{4\pi}{3}\Omega_0 \rho_c R_s^3.$$

In linear theory, the initial *rms* density contrast extrapolated to the present epoch ($z=0$) is:

$$\sigma_0(M) = \frac{D(z=0)}{D(z_i)}\sigma_i(M).$$

In this equation, $D$ is the growing mode of the linearized equation for the growth of perturbations and $z_i$ is the redshift corresponding to an early time at which the fluctuations in the universe were still linear. In the case of $\Omega_0 = 1$ and for a vanishing cosmological constant $\lambda_0 = 0$, $D \propto a$ with $a$ being the expansion factor. At $z=0$ and for $\lambda_0 = 0$, Peebles (1980) gives a good analytical fit for $f = d\ln D/d\ln a$: $f \sim \Omega_0^{0.6}$. Lahav *et al.* (1991) have extended the result to $z \neq 0$ and a non-zero cosmological constant case. However, in the case of our interest, a low density open universe, the expression can be analytically calculated (Weinberg 1972) and it is given in the Appendix. At very high redshifts, the behavior of the growth factor in the critical universe and in the open universe are similar. Due to the more rapid expansion of the scale factor, the growth of the perturbations at intermediate redshifts is much slower in the open universe than in the critical universe.

This implies a fundamental difference between open and critical models: the formation of structures is expected to occur much earlier in open models.

The exact condition for the non-linear collapse of a structure in a hierarchical theory is a rather complicated question. One generally uses the spherical model which greatly simplifies the problem: it appears from N-body simulations (see for instance Thomas & Couchman, 1992) as well as from analytical arguments (Bernardeau 1994) that it is a good description of the non-linear evolution of a perturbation. Within this model, we can calculate the critical overdensity $\delta_{c,0}(\Omega_0, z)$ that an object collapsing at redshift $z$ would have at redshift $z = 0$ in the linear regime. The derivation of $\delta_{c,0}(\Omega_0, z)$ for open universes is given in the Appendix.

Considerable attention has been devoted in the literature to the problem of the determination of the mass function. Most of these works were done in the case of the critical universe. However, due to the similarities in the set down problem, we can extrapolate the results to the low-$\Omega_0$ case. Using the spherical model, Press & Schechter (1974) proposed an estimation of the fraction of the mass of the universe that is in virialized objects. In the case of a gaussian random field, their mass distribution function is given by:

$$N(M,z)dM = -\sqrt{\frac{2}{\pi}}\frac{\overline{\rho}}{M}\frac{\delta_{c,0}(\Omega_0,z)}{\sigma_0(M)^2}\frac{d\sigma_0}{dM}\exp\left(-\frac{\delta_{c,0}(\Omega_0,z)^2}{2\sigma_0^2(M)}\right)dM.$$

Although the PS derivation contains some uncertainty, Blanchard *et al.* (1992) argued that it is physically motivated and likely to give a good approximation of the mass function. In practice, the use of the PS formalism is justified by its amazingly good fit to the mass functions found in numerical simulations. This result was first emphasized by Efstathiou *et al.* (1988) who compared the multiplicity function of collapsed structures in the case of the critical universe and for different values of $n$, to the PS mass function. From their simulations we can consider that the PS formula is valid for objects which contribute to less than $10^{-4}$ of the total density. In the case of galaxy clusters whose masses are of the order of $10^{15}M_\odot$, this density corresponds to a space density close to $10^{-7}h^{-1}(10^{15}M_\odot)^{-1}\mathrm{Mpc}^{-3}$. Recently, the PS formalism has also been compared to N-body simulations in the case of a low density, zero curvature model (White *et al.* 1993). The predictions of the theory agree quite well with the results of the simulations down to an abundance of $10^{-7}\mathrm{Mpc}^{-3}$ ($h = 0.5$). As $\lambda_0$ only plays a role in the growing rate of the fluctuations and does not affect the dynamics at $z = 0$, we take the above simulations that included a cosmological constant, for an additionnal justification of the PS formalism down to the same limits.

## 2.2. The $T - M$ and the $L_X - M$ relations

In order to test the mass function on galaxy cluster scale, we need to relate the cluster mass to some observed quantity like the X-ray temperature or the X-ray luminosity. The X-ray emission of galaxy clusters is produced by thermal bremsstrahlung from the hot, optically thin intracluster plasma which is shock heated during the infall onto the collapsing structure (Evrard 1990a, 1990b). It is usually assumed that the intracluster gas is isothermal and that it is in hydrostatic equilibrium within the potential of the cluster. For more details concerning the distribution of the gas under this hypothesis, see OBB. In this paper, we restrict ourselves to changes introduced by the low $\Omega_0$ case.

From the equation of the hydrostatic equilibrium, one can relate the X-ray temperature to the virial mass,

$$kT = \frac{G\mu m_p}{3\beta_{\text{fit}}} \frac{M_v}{R_v}. \tag{1}$$

In this equation, $\mu$ and $m_p$ are the mean molecular weight and the proton mass respectively and $\beta_{\text{fit}}$ is a parameter coming from the observed surface brightness profiles of clusters. The virial mass $M_v$ and the virial radius $R_v$ are quantities evaluated within a region where virialization has taken place. This regime is supposed to be achieved when the radius of the fluctuation is half of its maximum value. Assuming the spherical top-hat model, the density contrast $\Delta(\Omega_0, z) = \rho_v(z)/\overline{\rho}(z)$ of such a region can be derived at any redshift (cf. to the Appendix). The virial radius is then:

$$R_v = \left(\frac{4\pi\Omega_0 \rho_c \Delta}{3}\right)^{-1/3} M_v^{1/3}(1+z)^{-1}.$$

Using the typical value of $\beta_{\text{fit}}$, $\beta_{\text{fit}} \sim 0.6$ (Jones & Forman, 1984), and taking into account the incomplete thermalization of the gas which appears from the hydrodynamic simulations performed by Evrard (1990a, 1990b), the relation between the temperature and the virial mass becomes:

$$kT = 4\,\text{keV}\,\Omega_0^{1/3}(\Delta(\Omega_0, z)/178)^{1/3} M_{15}^{2/3}(1+z),$$

where $M_{15}$ is the cluster virial mass in units of $10^{15} M_\odot$. The incomplete thermalization represents an approximately 10% lowering in the normalization of the above relation. However, the shape of this relation is quite well reproduced by the simulations (Evrard, 1990a). It should be noticed that because of the frequency shift in an expanding universe, the observed or apparent temperature does not depend on redshift $T_{\text{app}} = T/(1+z)$.

It would be extremely usefull to derive a similar relation for the X–ray luminosity. However, the bolometric X–ray luminosity is given by:

$$L_X = \int_0^\infty \rho_{\text{gas}}^2(r) T^{1/2} 4\pi r^2 dr. \tag{2}$$

which depends on the gas density profile, and in particular, it is very sensitive to the core density which produces most of the X–ray luminosity. Since the formation of the X–ray core radius is determined by complicated competing processes (cooling flow, galactic feedback) which are not yet well understood, the $L_X - M$ relation is not easy to model. Indeed, a simple self–similar scaling (Kaiser, 1986) does not fit the observed local $L_X - T$ relation (OBB and references therein). Therefore, throughout this paper we use:

$$L_X = 3.3 \left(\frac{T_{\text{keV}}}{4}\right)^3 10^{44} \, \text{erg s}^{-1}, \tag{3}$$

which comes from the observations at $z = 0$ (Edge & Stewart, 1991). Since this expression represents the bolometric X–ray luminosity, we must correct it for the fraction

$$f_{\text{band}}(z) = \int_{E_1(1+z)}^{E_2(1+z)} dE \exp\left(-E/kT\right)/kT$$

which is collected in the interesting energy band $[E_1-E_2]$ (in this equation we have assumed that the Gaunt factor is equal to 1).

Concerning the evolution of the $L_X - T$ relation with redshift, Kaiser (1991) and Cavaliere *et al.* (1993) assert that clusters of a given temperature must have lower luminosities at higher redshifts in order to reproduce the negative evolution of the luminosity function (we shall return to this point beyond). However, Henry *et al.* (1994) claim that the evolution of cluster temperatures is moderate up to a redshift of 0.33 for clusters whose luminosity is close to $3.7 \times 10^{44} \text{erg s}^{-1}$. Therefore, unless we specify it, we assume that relation 3 does not depend on redshift. We will develop the implications of such an assumption within the last Section.

## 3. Constraints on $n$, $b$ and $\Omega_0$ from the temperature distribution function

In this Section, we constrain the free parameters of the model by fitting the theoretical temperature distribution function to the data at $z = 0$. In OBB we explain why we prefer to use the temperature distribution function rather than the luminosity function to constrain the power spectrum of the density fluctuations.

Up to now, there are two observed cluster temperature distribution functions which were derived from previously existing all–sky catalogues of X–ray surveys. These two distribution functions are those of Edge *et al.* (1990) and Henry & Arnaud (1991), which both are described in OBB. The range of temperature covered by the data goes from 2 to almost 14 keV. In the case $\Omega_0 = 0.2$, these temperatures correspond to linear scales of the order of $10h^{-1}$ to $25h^{-1}$Mpc. As a consequence, we can constrain both the normalization and the shape of the initial power spectrum on these scales.

The normalization of the power spectrum is conventionally specified through the variance, $\sigma_8^2$, of the mass overdensity within spheres of radius $8h^{-1}$Mpc. Since the fluctuations in the galaxy distribution give $\sigma_{\rm gal}(8h^{-1}{\rm Mpc}) \sim 1$ (Davis & Peebles, 1983), estimates of $\sigma_8$ are referred to as the bias parameter,

$$b = \frac{1}{\sigma_8}, \tag{4}$$

On the other hand, the mass within the normalization sphere depends on the value of $\Omega_0$. The free parameters of the temperature distribution function are then $n$ which mostly determines the shape of the function, and $\Omega_0$ and $b$ which both determine the characteristic temperature, ie. the temperature of a structure of mass $M_*$ such that $\sigma_0(M_*) \sim 1$. This characteristic temperature is:

$$T_* \sim 4.8\,\Omega_0^{2/3}\,b^{-\frac{4}{n+3}}\,{\rm keV}. \tag{5}$$

(We have neglected the factor $\Omega_0^{1/3}(\Delta/178)^{1/3}$, as it varies only slightly with $\Omega_0$. Actually, it is equal to 0.97 and 0.85 at $z = 0$, for $\Omega_0 = 0.8$ and 0.2 respectively). It is interesting to notice that this expression does not depend on the Hubble constant (see OBB), and therefore that the constraint we will derive are independent of its value.

In low–$\Omega_0$ models there is no motivation for a biased picture. However, there is some uncertainty in the observed value of $\sigma_{\rm gal}(8h^{-1}{\rm Mpc})$. This implies that the unbiased case could correspond to a value of $b$ in equation 4, somewhat different from one. We have then allowed $b$ to be also a free parameter. Because both $b$ and $\Omega_0$ enter in equation 5, there is a degeneracy between the two parameters. Therefore, in the set of three parameters fixing the shape and the normalization of the temperature distribution function, we have prefered to vary simultaneously either $(n, \Omega_0)$ or $(n, b)$. In the latter case, $\Omega_0$ is set to 0.2 which is the value obtained in usual density estimates (Peebles, 1986).

We have performed chi–square fitting of our model to each of the two existing observationally estimated temperature distribution functions. In the following, Edge *et al.*'s (1990) and Henry & Arnaud's (1991) temperature distributions are noted (ESFA) and (HA) respectively. The subscript 1 means that it was used in the originally published form and the subscript 2 signifies that we have added a point to the original data points.

This point represents A2163, the 14 keV cluster whose temperature was determined by Arnaud *et al.* (1992). The density at this point was calculated by assuming that there is one cluster of this temperature in the Abell survey (1958) which covers $\sim 1.53 \times 10^4$ deg$^2$. The Abell catalogue is complete only between $z \sim 0.02$ and $z \sim 0.2$ for clusters of richness class $R > 0$. However, it is likely that rich clusters at higher redshifts could have been included in the Abell catalogue. We have therefore assumed completeness out to $z \sim 0.3$. In all cases, the error bars represent the 90% confidence limits. In the case of HA and A2163, these errors bars correspond to Poisson statistics. We have then applied the same procedure than in OBB: the best-fit parameters are derived by chi-square minimization, contours are drawn in the two-dimensional parameter space and the uncertainties are obtained by checking by eye the goodness of fit of models lying in those contours (Fig. 1).

**Table 1.** Best–fit parameters

|         | $n$   | $\Omega_0$ | $b$  |
|---------|-------|------------|------|
| ESFA$_1$ | $-1.53$ | **0.20**   | 0.97 |
|         | $-1.55$ | 0.21       | **1**    |
| ESFA$_2$ | $-1.55$ | **0.20**   | 0.97 |
|         | $-1.56$ | 0.21       | **1**    |
| HA$_1$   | $-1.10$ | **0.20**   | 0.73 |
|         | $-1.39$ | 0.32       | **1**    |
| HA$_2$   | $-1.20$ | **0.20**   | 0.75 |
|         | $-1.42$ | 0.32       | **1**    |

In Table 1, and for each data set, we give the best-fit parameters in the case $\Omega_0 < 1$. The values given in thick line were fixed beforehand as explained above. For a given value of $\Omega_0$, and owing to the difference in the normalization of the two observed temperature distribution functions (a factor of two higher in the case of HA), the best-fit bias parameter is higher in the case of ESFA than in the case of HA.

In Fig. 1a, $\Omega_0$ is set to 0.2 and $(n, b)$ are left as free parameters. The contours of constant $\chi^2$ in the $n - b$ space are strongly elongated (Fig. 1a) which means that $\sigma_8$ is correlated to the power spectrum shape: this is due to the fact that the sphere which encloses the typical mass of a cluster has a radius which is of the order of $14h^{-1}$Mpc in the case $\Omega_0 = 0.2$. Therefore, contrary to the $\Omega_0 = 1$ case, clusters do not provide an unique normalisation at $8h^{-1}$Mpc.

In Fig. 1b, $b$ is set to 1 and $(n, \Omega_0)$ are the free parameters. It appears from this figure that the lower the value of $\Omega_0$ the lower the index of the power spectrum is, i.e. low $\Omega_0$ models require more power on large scale in order to produce the observed clusters.

From our analysis, we can now estimate the range of $\Omega_0$ which is consistent with the observed temperature distribution function at $z = 0$. Assuming that the *rms* fluctuation of galaxy density within spheres of radius $8h^{-1}$Mpc is equal to 1 and that the galaxy distribution is unbiased relative to that of the mass, we infer a value of $\Omega_0$ equal to $0.2 \pm 0.05$ from ESFA and equal to $0.3 \pm 0.1$ from HA. The higher value in the latter case is due to the higher normalization of the temperature distribution function as determined by Henry & Arnaud (1991). Relaxing the assumption that $b$ is equal to 1, we fix the value of $\Omega_0$. This approach is preferable because of the degeneracy between the two parameters which both determine the degree of clustering of the mass. In the case $\Omega_0 = 0.2$ which is the dynamical estimate of the density parameter, a value of $b$ equal to 1 is compatible with ESFA whereas the best–fit bias parameter is close to 0.7 from HA. Recently, also using the PS formalism, White *et al.* (1993) computed the value of $\sigma_8$ necessary to reproduce the observed density of Abell clusters in flat universes with different values of $\Omega_0$. They assumed an abundance of $5 \times 10^{-7}$Mpc$^{-3}$ for clusters with masses equal to $1.1 \times 10^{15} M_\odot$ within an Abell radius and equal to $1.3 \times 10^{15} M_\odot$ within the virial radius (for $h = 0.5$). From Edge *et al.*'s (1990) and Henry & Arnaud's (1991) cumulative temperature distribution functions, they took a mean value of 3.6 keV at this abundance. White *et al.* (1993) did not attempt to extract information on the shape of the power spectrum from clusters, so we can only compare our results to the value of the bias parameter they derived. In the case $\Omega_0 = 0.2$ they found that the value of $b$ has to be in the range 0.6–0.8 in order to produce clusters with sufficient masses within an Abell radius at the standard abundance they assumed. This result is in agreement with the value of the bias parameter we obtain by fitting the theoretical model to HA. On the contrary, due to the lower normalization of ESFA, we find that an unbiased model is compatible with their observations. Concerning the initial power spectrum index we derive from the observed temperature distribution function, we find a value of $n$ close to $-1.5$ from ESFA, and between $-1.4$ and $-1.1$ from HA. The difference is due to the scaling of the characteristic mass corresponding to the exponential cut–off in the PS mass function, which depends on both $\Omega_0$ and $b$, and is not due to the intrinsic difference in the shape of the two observed temperature distribution functions. It is clear that because of the degeneracy between $b$ and $\Omega_0$, the temperature distribution function of X–ray clusters at zero redshift does not allow to distinguish the open model from critical models. Nevertheless, our analysis has revealed the interesting fact that in the case of an unbiased open model, the temperature distribution function implies a value of $\Omega_0$ in the range $0.15 - 0.4$: unbiased models with $\Omega_0 \leq 0.15$ cannot reproduce the temperature distribution function. It should be noticed that although this determination is formally independent of dynamical estimates on cluster scale, the results are very similar.

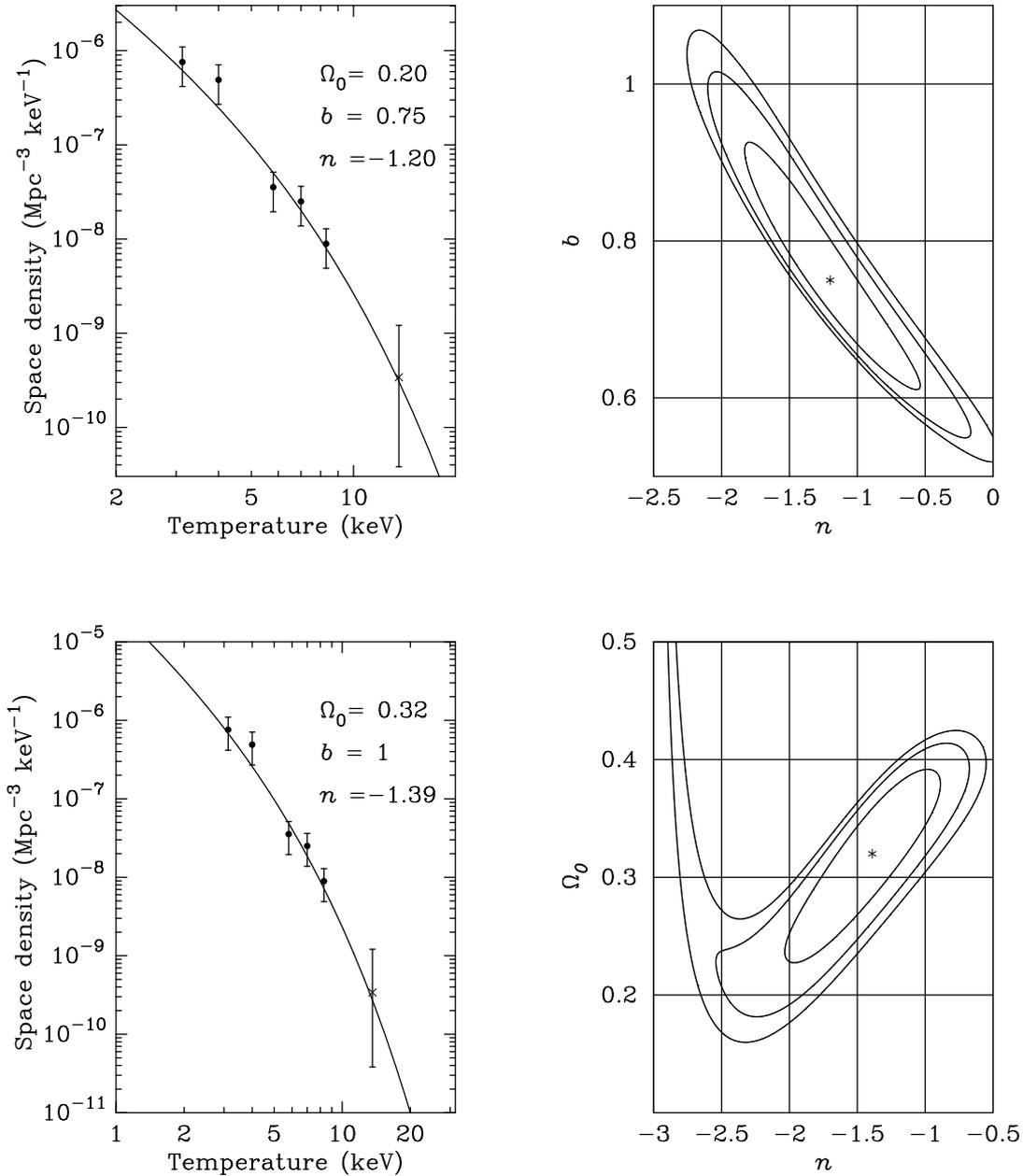

**Fig. 1.** The left hand side column shows Henry & Arnaud's (1991) temperature distribution function and the best–fit theoretical function. The right hand side column shows confidence region ellipses corresponding to $\Delta\chi^2 = 2.30, 4.61, 6.17$. These contours correspond to 68.3%, 90% and 95.4% respectively, for normally distributed data. From top to bottom: (a) The value of $\Omega_0$ is set to 0.2. (b) The value of $b$ is set to 1

## 4. What can we learn from the redshift distribution ?

As we have already mentioned, structure formation occurs at different rate according to the present mean density of the universe. Evolutionary properties are therefore expected to differ with $\Omega_0$. Such a difference is expected to appear from various observations, and in particular from the redshift distribution that we will now investigate in detail. In order to put in light this difference, we will compare the predictions given by two different models which fit equally well the data at zero redshift. These models are those which give the best-fit to HA in the case $\Omega_0 = 1$ ($n = -1.85$ and $b = 1.65$, see OBB) and in the case $\Omega_0 = 0.2$ (see Table 1). Hereafter, these two models are referred to as the critical model and the open ($\Omega_0 = 0.2$) model respectively.

### 4.1. Insensitivity to the spectrum

The evolution of the X-ray cluster population has been discussed by Kaiser (1986) in his pioneering work. In the case $\Omega_0 = 1$ and using scaling arguments, he concluded that the density evolution of X-ray selected clusters should reflect the index of the initial power spectrum. This can be seen in the following way. For power-law initial fluctuation spectra, Kaiser (1986) showed that the evolution of the statistical properties of clusters in the critical universe is self-similar. This means that, by applying an appropriate scaling to some cluster property, for instance the temperature, one can recover the distribution function at any epoch from the observed distribution function at $z = 0$. The only physical scale in the problem is the characteristic mass which is going non-linear and which varies as $M_{\rm NL} \propto (1+z)^{-6/(n+3)}$. The higher the value of $n$, the slower the decrease of the characteristic mass. Kaiser (1986) used this argument to claim that the evolution of the comoving density of galaxy clusters depends on the value of the initial power spectrum index. This has been illustrated by Pierre (1991), who applying the above scaling to the local ($z = 0$) temperature distribution function and converting it into a luminosity function, found a conspicuous difference in the redshift distribution between models with values $n = -1$ and $n = -2$ of the power spectrum index. However, the above arguments contain a loophole and the implicit assumptions which have been made are not self-consistent. Let us clarify this point by explicitly using the general form of the mass function. As we mentioned in Section 1, the mass function can be written as:

$$N(M, z = 0) = \frac{\overline{\rho}}{M^2} \nu_0 \frac{d \log \sigma_0}{d \log M} F(\nu_0),$$

$\nu_0$ being $\delta_{c,0}(z = 0)/\sigma_0(M)$. Numerical simulations suggest that the function $F$ is well fitted by $\sqrt{2/\pi} \exp(-\nu^2/2)$ ($\nu = \delta_{c,0}(z)/\sigma_0(M)$), independently of the power spectrum. Therefore, if the present-day number is fitted on some mass scale, the amplitude of the fluctuations, $\sigma_0(M)$ is fixed in a way which does not depend on the power spectrum (neglecting the change in the coefficient $d \log \sigma_0 / d \log M$). Since $\delta_{c,0}(z) = \delta_{c,0}(z = 0)(1 + z)$ in the critical universe, the number density can be evaluated at any redshift:

$$N(M, z) = N(M, z = 0) \frac{F(\nu_0(1 + z))}{F(\nu_0)} (1 + z).$$

This clearly does not depend on the power spectrum. For theoretical models normalized to the scale $8h^{-1}$Mpc which corresponds to clusters of $\sqrt[3]{\Omega_0} 4.5$ keV, the redshift distribution of $\sqrt[3]{\Omega_0} 4.5$ keV apparent temperature clusters (the apparent temperature corresponds to the same mass whatever the redshift) will be the same for different spectra. If the selection procedure is based on some other quantity (like the luminosity), the same argument applies provided that the relation of the quantity to the mass is independent of the spectrum. Actually, only measurements on two different mass scales can lead to information on the power spectrum. When one compares the redshift distribution for different masses, there is some difference because one is sampling different parts of the function $F$, but this difference is weak. Furthermore, such difference will also occur by changing the normalization. This is illustrated in Fig. 2. Of course, the total number of objects is not the same for the two power spectra, reflecting the fact that the local properties are different. We conclude that the redshift distribution does not carry useful information on the power spectrum. On the contrary, the redshift distribution depends on the cosmological parameters ($\Omega_0$, $\lambda_0$). In original Kaiser's scaling argument (1986) it was assumed that whatever the power spectrum is, the local observed distribution function is reproduced. This is of course not true, since different spectra should lead to different distribution functions which cannot fit simultaneously all the observed local distributions.

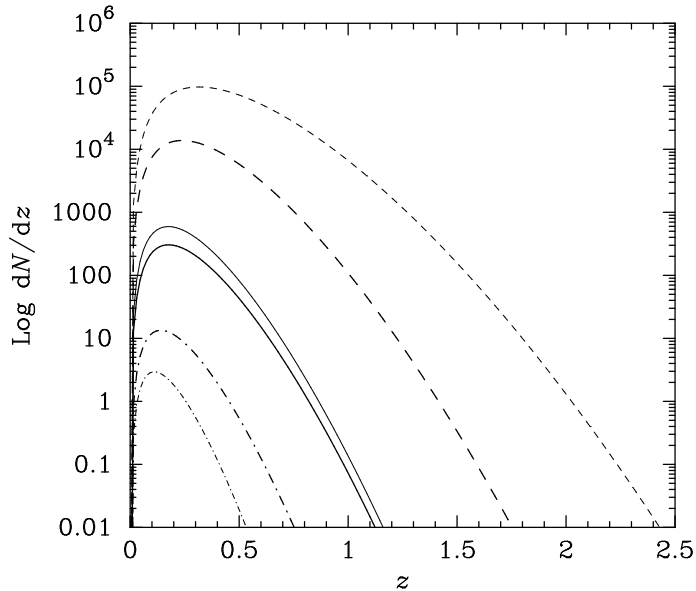

**Fig. 2.** The redshift distribution of clusters in all the sky, in the case of the critical universe. Two models are represented: a model with the spectral index, $n$, equal to $-2$ (thick lines), and a model with $n$ equal to $-1$ (thin lines). In both cases, the bias parameter $b$ is equal to 1.7. The different lines correspond to different apparent temperatures: $T = 2\,\mathrm{keV}$ (dashed line), $T = 4.5\,\mathrm{keV}$ (solid line) and $T = 8\,\mathrm{keV}$ (dotted–dashed line)

*4.2. The redshift distribution of galaxy clusters according to the value of $\Omega_0$*

In this Section we investigate in more detail the dependence on $\Omega_0$ of the characteristics of the X–ray cluster population at high redshifts. We mainly focus on whether this could provide an useful test of the mean density of the universe and distinguish between open and critical density models. In particular, we predict the redshift distribution of galaxy clusters expected to be observed by the ROSAT satellite in both models. The ROSAT satellite performed the first all-sky survey using an imaging X–ray telescope. It covers the energy band between 0.1 and 2.4 keV and its detection limit lies close to $10^{-12}\,\mathrm{erg\,cm^{-2}\,s^{-1}}$ for most of the sky (Böhringer *et al.*, 1992). Around the ecliptic poles and for a region which spans roughly a 10 deg radius area, the flux limit is close to $10^{-13}\,\mathrm{erg\,cm^{-2}\,s^{-1}}$. Nearly 50000 sources have already been detected within the all–sky survey. A first identification of the galaxy clusters within the southern sky is based on the comparison of the source list with an optical galaxy catalogue, as well as on the extent of the X–ray sources (Guzzo *et al.*, 1995). However, such criteria will probably introduce a bias towards nearby clusters since optical catalogues are complete only at low redshifts

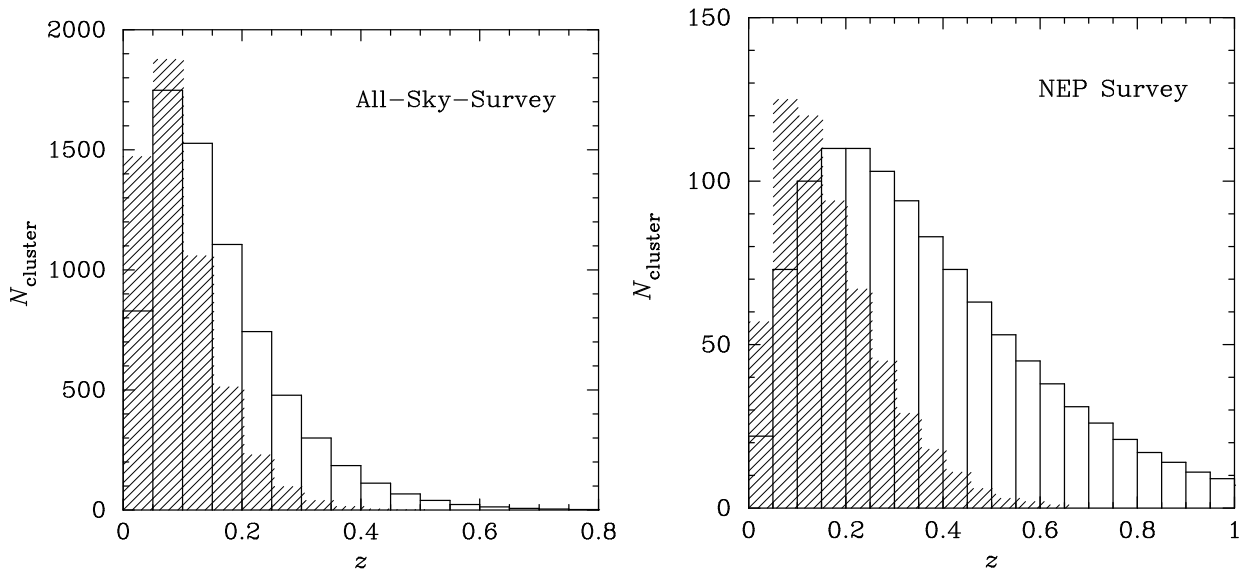

**Fig. 3.** Expected redshift distribution of clusters in the All–Sky survey of the ROSAT satellite in the case of the critical universe (shaded area) and in the case of the open universe (thick solid line). The critical model corresponds to $n = -1.8$ and $b = 1.6$, and the open model corresponds to $\Omega_0 = 0.2$, $n = -1.2$ and $b = 0.7$. In both cases, the observed local $L_X - T$ relation is used (3), with the assumption that it does not vary with redshift. (a) The All–Sky survey : the flux limit is close to $10^{-12}\,\mathrm{erg\,s^{-1}\,cm^{-2}}$. (b) The North Ecliptic Pole survey : the flux limit is close to $10^{-13}\,\mathrm{erg\,s^{-1}\,cm^{-2}}$

and most of the clusters have angular core radii which are below the resolution of the instrument and are therefore classified as point sources and missed as clusters.

Let us now examine the expected redshift distribution of the $\Omega_0 = 1$ and $\Omega_0 = 0.2$ models within the above described flux limited surveys, in the case where the $L_X - T$ correlation does not undergo any evolution. In this case, the luminosity is related to the mass according to the following scaling:

$$L_X \propto M^2 (1+z)^3$$

Figure 3 shows the expected number of clusters detected per redshift bin in the All–Sky survey (Fig. 3a) and in the north ecliptic pole region (NEP) (Fig. 3b) for both the critical (shaded areas) and the open (solid lines) models. The total number of clusters at the flux limit of the All–Sky survey is different according to the model: while 7200 clusters are expected in the open case, 5300 are expected in the critical case. At low redshifts, below $z \sim 0.15 - 0.2$, the cluster distribution is identical in both models. This result was expected since the models were normalized to match the local observed temperature distribution function. In the case of the critical model, strong evolution appears at a redshift as low as 0.2 resulting in a net decrease, and there is no cluster expected beyond $z \sim 0.4$. In the open model, clusters form earlier and their distribution function remains almost constant with redshift. Consequently, the expected number does not decrease as fast as in the critical density case: more than 100 clusters lie beyond redshift 0.4. However, the high redshift clusters still represent a small fraction of the total number of clusters expected at this flux limit: assuming the absence of any bias in the selection procedure, optical spectroscopic follow up of at least 10% of the survey should be achieved in order to probe the difference significantly. In Fig. 4, we show the expected redshift distribution in the NEP for the two models we have investigated. We recall that the flux limit in the NEP region is close to $10^{-13} \mathrm{erg\,cm^{-2}\,s^{-1}}$. At this sensitivity, clusters are actually probed up to higher redshifts and the redshift distribution is significantly different. This difference mainly occurs because of the larger number of high temperature clusters at high redshifts in the case of the open model. In this model, half of the sources are expected beyond $z \sim 0.32$, while the median redshift is only $z = 0.14$ in the case of the critical model. Although the total number of expected clusters is smaller in the NEP region than in the All–Sky survey (the NEP survey is expected to contain less than 15% of the All–Sky survey), our analysis reveals the fact that the redshift distribution of clusters in the NEP offers a better way to differenciate a low density universe from the critical universe. Nevertheless, hotter clusters are more luminous and their detectability should be less affected by the survey flux limit. The redshift distribution of the hottest clusters within a flux limited sample can therefore provide the adequate information. This is illustrated in Fig. 5 showing the redshift distribution of clusters whose apparent temperature is 6 keV. The redshift distribution is given for various flux limits, in the case of the critical model. Clearly when the flux limit is low enough, the redshift distribution should depend only on the density parameter. Therefore, the temperature information allows one to overcome the high flux limit problem. Let us now examine this point quantitatively. As an illustration, we focus our predictions for samples whose characteristics are similar to those of ROSAT. For clusters which apparent temperature is higher than 2.4 keV (although the temperature information cannot be obtained over the All–Sky survey, this temperature limit ensures one that the cutoff of these clusters X–ray spectra is above 2.4 keV which is the higher energy limit of the ROSAT window), the redshift distribution is almost identical for the two models as we can see in Fig. 3a, and is not significantly different than in the case where only a flux selection is used. This is due to the fact that the distribution is dominated by clusters whose temperature is close to 2.4 keV: again, these clusters, owing

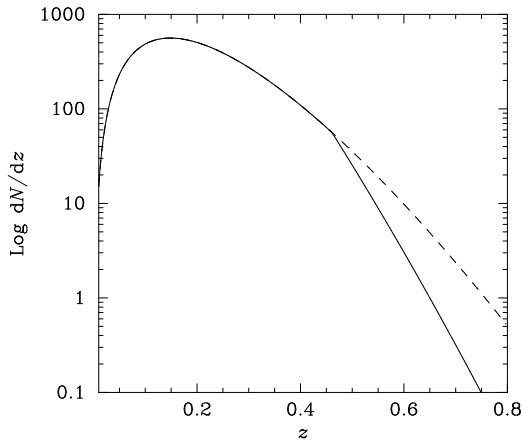

**Fig. 4.** The differential redshift distribution of clusters in all the sky in the case of the critical model ($n = -1.8$ and $b = 1.6$). The different lines correspond to different flux limits: $10^{-13} \mathrm{erg\,cm^{-2}s^{-1}}$ (solid line) and $10^{-12} \mathrm{erg\,cm^{-2}s^{-1}}$ (dashed line)

to their low luminosity are detected essentially at low redshifts. However, there are more clusters of this apparent temperature range in the open model than in the critical model (5411 and 2628 respectively). In Fig. 3b, we give the redshift distribution of clusters with apparent temperatures higher than 5 keV. The median redshift is $\sim 0.3$ in the case of the open model and $\sim 0.2$ in the case of the critical model. Clusters of $T_{\mathrm{app}} > 5$ keV are enough luminous to be detected even at high redshifts: in the case of the open model there are almost 600 clusters which could be detected between $z = 0.25$ and $z = 0.35$. Only 100 are expected in the case of the critical model. This large difference can easily be detected with few optical spectroscopic measurements. Of course, the difference in the redshift distribution increases with temperature: in Fig. 3c, we show the case of clusters with $T_{\mathrm{app}} > 8$ keV. These objects lie on the exponential cutoff of the temperature distribution function and the median redshifts are noticeably different for the two models: 0.37 and 0.16 for the open model and the critical model respectively. The redshift distribution in this case is a very powerful test, even at a sensitivity limit of $10^{-12} \mathrm{erg\,cm^{-2}s^{-1}}$. However, a complete optical identification of the ROSAT sources is needed – at least in a given region of the sky – in order to make the test feasible. Moreover, reliable data on the temperature are also necessary. Although this information can probably not be obtained over the All–Sky survey, it is conceivable that a satellite like AXAF or XMM will make temperature measurements on a subsample of ROSAT selected clusters in order to perform the test we propose.

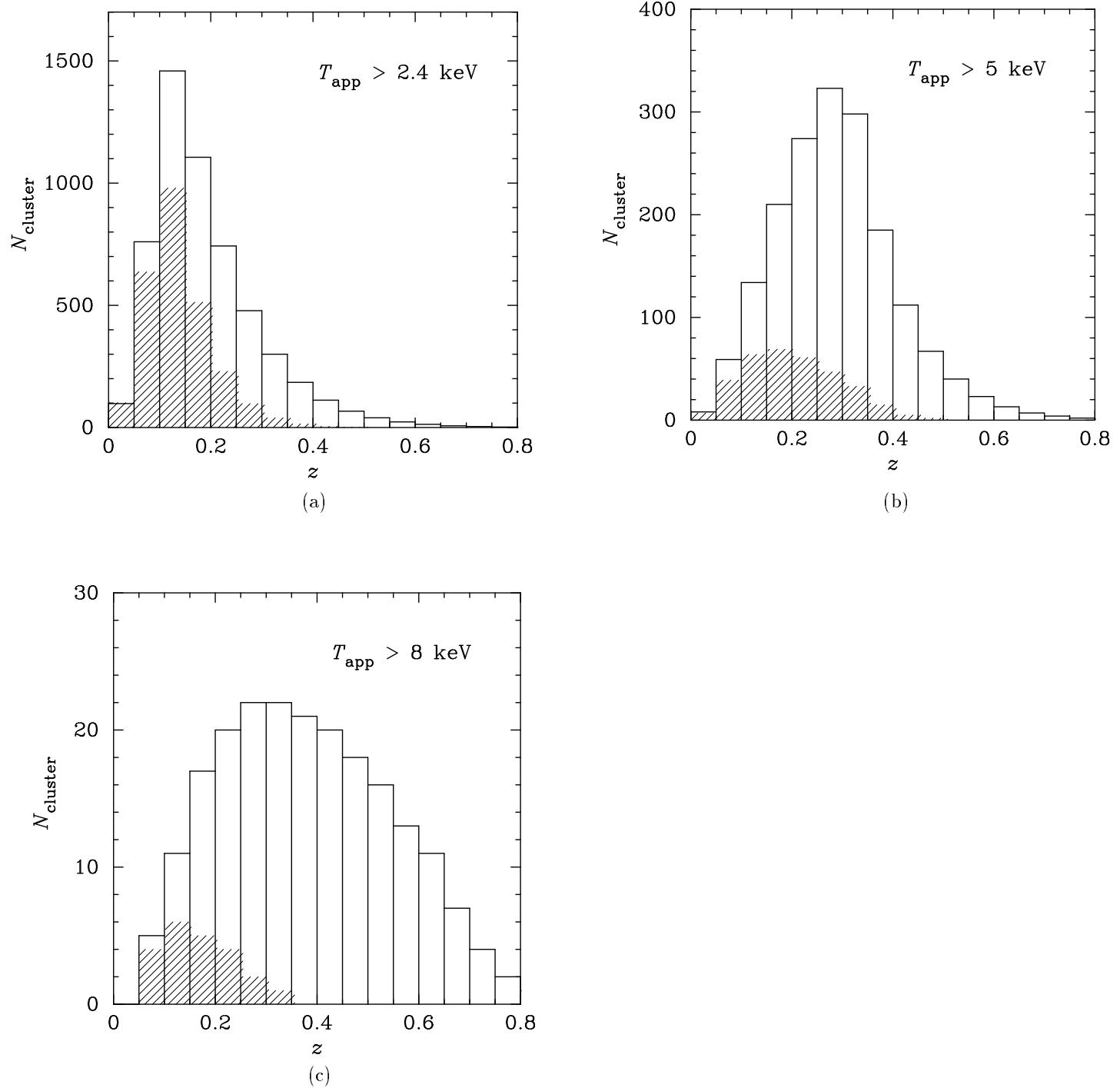

**Fig. 5.** Expected redshift distribution of clusters in the All–Sky survey of the ROSAT satellite in the case of the critical universe (shaded area) and in the case of the open universe (thick solid line). Models are as in Fig. 3. (a) Clusters whose apparent temperature is higher than 2.4 keV. (b) Clusters whose apparent temperature is higher than 5 keV. (c) Clusters which apparent temperature is higher than 8 keV

This analysis shows that the temperature information is critical to actually access to the density parameter, at least for a high flux limited survey. This is because the redshift difference comes essentially through the hottest clusters: while the redshift distribution of low temperature clusters ($T \leq 2.5$ keV) is almost identical in both models, the difference is unambigous and significant for clusters with $T > 5$ keV. Since the difference increases as the flux limit lowers, the knowledge of the temperature does not seem crucial in order to distinguish the two models within a low flux limited survey. However, this last result is no more true if we consider the possible evolution of the $L_X - T$ relation. Indeed, the redshift distribution within a flux limited survey could be affected by evolution in luminosity. For example, one can imagine that a negative evolution of the $L_X - T$ relation within the $\Omega_0 = 0.2$ model could mimic the NEP redshift distribution of the non-evolving $\Omega_0 = 1$ model. No definitive conclusion could be therefore drawn on the single basis of the redshift distribution and the temperature information is necessary to conclude. On the other hand, the possible evolution of the $L_X - T$ correlation is comparable to the effect of a high flux limited survey, and does not affect the redshift distribution of the hottest cluster. It is therefore clear that the determination of the density parameter can be achieved by the redshift distribution test, provided that high redshift temperature measurements exist. The possible evolution increases the complexity of the analysis, but does not represent a intrinsic limitation of the method.

*4.3. Comparison with existing data*

In the last years there have been several claims for evolution in the X-ray cluster population. Edge *et al.* (1990) first noticed evolution in the luminosity function. They derived source counts for two subsamples of their data: one with high luminosity ($L_X > 8 \times 10^{44}$ erg s$^{-1}$) clusters and the other with low luminosity clusters. They explained the deficit at low fluxes of the high luminosity subsample by a lack of high luminosity clusters at redshifts greater than 0.1. At the same time Gioia *et al.* (1990) using 67 X-ray selected clusters from the *Einstein Observatory* Extended Medium Sensivity Survey (EMSS) derived the luminosity function in three redshift shells: $0.14 \leq z < 0.20$, $0.20 \leq z < 0.30$ and $0.30 \leq z < 0.60$. They found significantly steeper slopes in the high redshift shells compared to the low redshift shell. More recently, to test the evolution of the X-ray luminosity function at higher redshifts, Castander *et al.* (1994) analyzed two deep ROSAT pointings containing distant optically-selected clusters of galaxies. Among the five optically rich cluster candidates they selected, they detected X-ray emission only from two clusters. Castander *et al.* (1994) claimed, that if the probability distribution

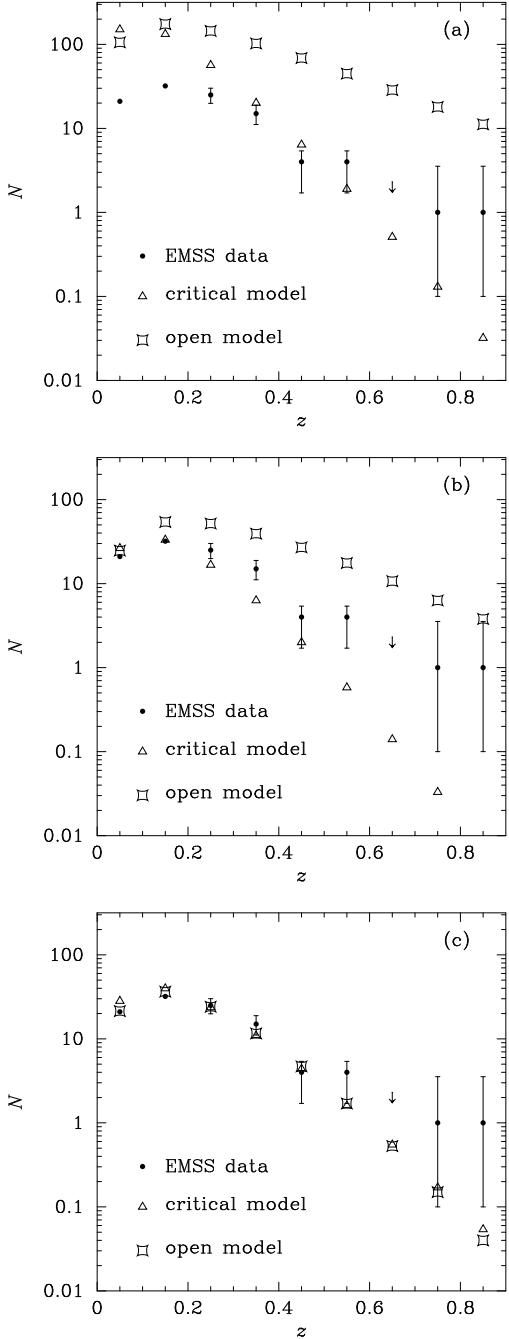

**Fig. 6.** The redshift distribution of clusters, taking into account the sky coverage for different sensitivities in the EMSS. The squares are for the open model and the triangles are for the critical model. The points represent the redshift distribution as given by Gioia & Luppino (1994). The errors are from Poisson statistics. The upper limit at $z = 0.65$ accounts for the fact that no cluster has been observed in the redshift bin centered at this point. (a) The redshift distribution in the case where the bias due to the cluster extension is neglected. (b) The redshift distribution including in each redshift bin the mean correction factor for the flux. (c) The redshift distribution of galaxy clusters in the open model assuming a negative evolution of the $L_X - T$ relation (see text)

that a cluster is selected is assumed to be equal to one in the redshift range $0.5 < z < 0.9$, then their survey volume is $7.4 \times 10^6 \mathrm{Mpc}^3$, and the luminosity function they derived is in agreement with the decline in the comoving density of clusters with redshift which has been already observed. Using similar search methods, Nichol *et al.* (1994) reached the same conclusions. Moreover, recent analysis of the EMSS data (Luppino & Gioia, 1995) seems to indicate that the density derived from high redshift X-ray selected clusters is in agreement with the density derived by Castander *et al.* (1994) and Nichol *et al.*'s results (1994). In its own, such an evolution favors critical models since no significant evolution of cluster properties is expected in open models.

Recently, Gioia & Luppino (1994) have given the redshift histogram for the entire EMSS cluster sample. We can then compare to these data the expected X-ray cluster redshift distribution in the critical model and in the open model respectively and apply the test we propose. Indeed, this sample should be very well adapted to our test since it is entirely X-ray selected and since almost all the sources in the survey have been identified. The main limitation comes from the fact that the temperature information is not yet available. An other difficulty comes from the extended nature of clusters which prevents the survey to be flux limited for these objects. Indeed, clusters with core radii more extended than the detection cell (2.4'×2.4') are likely to be missed and there is a strong bias at redshifts smaller than $z = 0.17$ for clusters with core radii greater than 300 kpc. Taking into account the sky coverage for different sensitivities in the EMSS, but neglecting the bias due to the extended nature of clusters, we give in Fig. 6a the theoretical redshift distribution corresponding to the survey. Low redshift bins are not well reproduced by the theoretical predictions. Since our models are normalized to low redshift data, this could not mean that the models are to be rejected, but rather illustrates the effect of the selection function. In order to account for this, we have estimated in each redshift bin the mean correction factor for the flux: this correction factor corresponds to the mean ratio of the actual cluster flux over its flux in the detection cell. The former quantity is given by Gioia & Luppino (1994) for each cluster in the EMSS sample. With the selection function modeled in this way, the redshift distributions are strongly modified at low $z$. This is illustrated in Fig. 6b. As one can see, the agreement between the theoretical predictions and the observations is quite good. It is likely therefore that our modeling of the selection function is reliable. In addition, since the correction factor is greater at low redshifts, we are confident that our correction for the higher redshift clusters is secure. From these figures it appears that the open model produces a large excess between $z = 0.2$ and $z = 0.5$, within the redshift interval which contains the bulk of the expected clusters.

There are 76 clusters detected within this redshift range, whereas 172 are predicted in the case of the open model. On the contrary, the $\Omega_0 = 1$ model fits the data rather well, although at high redshift, it slightly underestimates the total expected number.

These results were derived assuming no evolution of the $L_X - T$ relation, since as we have already mentioned, there is no sign of significant evolution of this relation out to a redshift of $z = 0.33$ (Henry et al., 1994). However, due to the small photon number of each cluster within Henry *et al.*'s sample (1994), only the average temperature of clusters was determined. Moreover, the highest redshift bin of their sample is $0.3 \leq z < 0.43$ whereas the redshift distribution of the EMSS sample extends out $z \sim 0.9$. Therefore, there is still room for possible evolution. We have then investigated possible implications of such an an evolution by modeling the $L_X - T$ relation at high redshifts by a power law with a shape identical to the shape measured at low redshift but with an evolving normalization:

$$L_X = c\,(1+z)^\beta T^3 \qquad (6)$$

where $c$ is the normalisation of the $L_X - T$ relation at $z = 0$ (Eq. 3) and $\beta$ is a free parameter. A chi-square fitting of our models to the EMSS cluster redshift distribution gives $\beta = 1^{+0.38}_{-0.63}$ and $\beta = -2.3^{+0.35}_{-0.3}$ for the critical and the open universe respectively (the given errors correspond to the 90% confidence intervals). These parameters were used in Fig. 6c. Obviously, these evolutionary laws allow one to fit the observed distribution and no information can be extracted on the density parameter anymore. In the same way, the X-ray cluster number counts and the contribution to the X-ray background will be identical in open and critical models (OBB).

Let us now investigate whether temperature information of high redshifts clusters would allow us to recover this information. For this, we considered the 8 EMSS clusters lying in the redshift range $0.4 \leq z < 0.6$. Taking into account the incertainties in the X-ray temperature measurements and the intrinsic dispersion of the $L_X - T$ correlation measured at $z = 0$, would the measurement of these cluster temperatures allow us to distinguish between the critical and the open universe? To answer this question we decided to use a simple bootstrap approach. Although the uncertainties are sometimes quite large, the X-ray temperature information is available for a substantial number of nearby clusters. In order to handle an uniform sample of good quality, we restricted ourselves to clusters for which the temperature were measured with the Ginga satellite, 27 in total (see references quoted in Fig. 2 of Arnaud 1994) The uncertainty in the determination of the $L_X - T$ correlation within this sample comes mainly from the intrinsic dispersion of the data points. We wanted to estimate the uncertainty in the normalisation of the $L_X - T$ correlation at high redhifts, in the case where only a small number of X–ray temperature measurements were made. This were done by the bootstrap resampling technic. Using this sample at low redshift, we created 2 synthetic catalogues at $z = 0.49$ (this corresponds to the mean redshift of the 8 EMSS cluster we considered): we modified the luminosity of each cluster according to the evolution needed to fit the redshift distribution for the critical and the open model respectively. Within these 2 new catalogues, we picked up all the clusters whose fluxes would have been within $3 - 7 \times 10^{-13}$ erg cm$^{-2}$s$^{-1}$ and $1 - 1.6 \times 10^{-12}$ erg cm$^{-2}$s$^{-1}$ at $z = 0.49$. These values are the fluxes of the $0.4 \leq z < 0.6$ EMSS clusters. There are 10 clusters corresponding to this criterion within each of the 2 synthetic samples. From each of these new reduced samples, we draw 10000 times a subsample of 8 clusters (we use the bootstrap technic, so each cluster could be drawn from the sample as many times as it happens), and we fit the normalisation of the data points by assuming that the shape of the relation is the shape measured at $z = 0$ within the original sample. From our simulations we notice that the individual computed incertaintities are 10 times lower than the dispersion resulting from the resampling. This confirms that the uncertainty comes mainly from the intrinsic dispersion in the $L_X - T$ relation. In practice, even with accurate temperature measurements, about ten clusters are necessary to infer a robust conclusion on the normalisation of the $L_X - T$ correlation. From our simulations, we estimated that the normalisation is equal to $1.9^{+1.5}_{-0.9} \times 10^{-2}$ and $5.1^{+4}_{-2.4} \times 10^{-3}$ at 90% confidence level, for the critical and the open model respectively. It appears that these two intervals do not recover each other, and therefore temperature measurement of about ten $z \sim 0.5$ clusters would allow to easily distinguish between the two models.

The evolution of the temperature distribution of X–ray clusters can clearly allow the determination of the value of the density parameter of the universe. A direct observational proof of this quantity is still far from being available, and therefore the application of our new test in its complete version is still prospective. We have shown that the redshift distribution of X–ray selected clusters from the EMSS survey favors a high value of the density parameter. Because of possible evolution of the $L_X - T$ relation this is however a quite uncertain indication. This is illustrated by our analysis in which we allowed for a possible evolution of the temperature–luminosity relation (Eq. 6). In such a case, the two models reproduce equally well the data, provided that the evolutionary law is adequately chosen. It is interesting to notice that the few observed high $z$ clusters are not well fitted by any of the models. This could be the indication that these clusters are spurious in some way. Quite obviously, as we have already emphasized, in a flux limited survey, possible unknown evolution of X–ray clusters prevent any confident conclusion on the density of the universe to be drawn from their past abundance: in order to carry the test we propose it is necessary to have the temperature information. Now, taking into account the information provided by the EMSS survey, it is only necessary to determine the possible evolution of the $L_X - T$ relation: in a low density universe clusters of a given temperature should be substantially fainter in the past. Using a bootstrap procedure, we have investigated the possibility of distinguishing the two models by this method: we show that the difference can actually be probed with around ten clusters for which temperature is measured with a realistic accuracy. This demonstrates the possibility of achieving our test with ASCA or XMM temperature measurments of a limited number of clusters.

## 5. Conclusion

The investigation of cluster properties within galaxy formation theories can provide fundamental informations on various aspects of the models. OBB have shown that a reliable modeling of X–ray clusters can be built and therefore, that the inferred constraints are robust. As other authors, they investigated the consequences of this modeling for flat models and found, in agreement with other analysis, that the spectrum on clusters scales ($\sim 8h^{-1}$Mpc) has to be flatter than that given by the standard cold dark matter model (see for instance Peacock and Dodds, 1994). In the present paper we have carried out a similar analysis in a low density universe. Because of the degeneracy between the amplitude of the fluctuations and $\Omega_0$, the local data does not allow one to distinguish between an open model and the critical universe. However, it provides an interesting result: in the unbiased case, though the derivation is drawn from a different analysis, the estimate of the mean density of the universe is consistent with dynamical estimates and $\Omega_0$ has to be greater than 0.15. In the second part of this paper we have analysed a new way to estimate the mean density of the universe. As structure formation occurs earlier in a low density universe than in the critical case, the evolutionary properties are different and this leads to important observational differences: we have shown that the redshift distribution of clusters selected on the basis of their apparent temperature is primarily sensible to the density parameter. Since we have also shown that this distribution is almost completely independent of the power spectrum, this test might provide an unambiguous determination of $\Omega_0$.

We have attempted to apply our test to the Einstein satellite X–ray survey. As the temperature information of these clusters is not available, our test cannot be fully carried out. Still, assuming a non-evolving relation between luminosities and temperatures at high redshifts, this first piece of information seems to favor an Einstein–de Sitter universe. However, a strong negative evolution of the luminosity–temperature relation might mimic the observed redshift distribution in a low density universe. Since present day data are clearly inconsistent with the simplest self-similar model, theoretical modeling should be taken with some caution. Indeed, it seems difficult to solve this problem from the theoretical point of view: the self-similar model implies that the luminosity, for a given temperature should be higher at higher redshifts, but scaling arguments do not allow to properly reproduce present day data. On the other hand, as structure formation occurs early in low density model and late evolution is a rather passive one, mainly due to the dilution of the universe, we believe that a strong negative evolution is rather unlikely. This gives support to the idea that the observed redshift distribution is not in agreement with an open model. However, such a conclusion can only be considered as tentative. Important progress on this question is expected from detailed observations of high redshift clusters: from our bootstrap analysis we have shown that temperature measurements of about ten high redshift ($z \sim 0.5$) clusters would allow us to fix the normalization of the $L_X - T$ correlation, and thus to distinguish between an open model and the flat model. Consequently, very sensitive instruments like AXAF and XMM could give a definitive answer to the question concerning the mean density of the universe.

**Appendix**

In this appendix, we give a simple derivation in the case of a low density universe, of the amplitude of the initial fluctuations that collapse at some redshift, as well as the density contrast of the virialized objects. The spherical collapse in an expanding universe was first studied by Lemaître (1933). Gunn & Gott (1972) did cluster formation analysis through this model and the modern version of the spherical collapse can be found in Peebles (1980). Here, we derive in an elegant way the quantities that are used in this paper. Although these expressions were used by Oukbir & Blanchard, they were not given. The expression for the threshold density contrast of the non-linear collapse, though derived in a different way, has also been given by Lacey & Cole (1993) (see their appendix).

*1. The density threshold*

At high redshifts, the amplitude of the density fluctuations are small and the perturbations grow according to linear theory: $\delta(\mathbf{x}, t) \propto D(t)$. In the case of a low density open model, the growth factor $D(t)$ is given by (Weinberg 1972):

$$D = -\frac{3\psi \sinh \psi}{(\cosh \psi - 1)^2} + \frac{5 + \cosh \psi}{\cosh \psi - 1}. \tag{7}$$

In this equation, $\psi$ is the development angle defined as,

$$\cosh \psi - 1 = \frac{2(1 - \Omega_0)}{\Omega_0} \frac{a(t)}{a(t_0)}, \tag{8}$$

with $a(t)$ being the expansion factor of the universe. When the amplitude of $\delta(\mathbf{x}, t)$ approaches unity, the overdense region decouples from the Hubble expansion, turns around, collapses and virializes. A simple analytic model allows one to calculate the critical overdensity corresponding to the redshift $z_v$ at which a region virializes. A useful quantity is the critical linear overdensity extrapolated to the present day: $\delta_{c,0}(z)$.

One considers a spherical overdensity within a homogeneous universe. In the case where there is no shell crossing, Birkhoff's theorem allows one to treat the evolution of the overdensity as if it were an unperturbed region of the universe with average density $\widetilde{\rho} = \widetilde{\Omega} \widetilde{\rho}_c$. Since $\widetilde{\Omega} > 1$, the parameteric equations of motion of the overdense shell are:

$$\frac{R(t)}{R(t_0)} = \frac{\widetilde{\Omega}(1 - \cos \theta)}{2(\widetilde{\Omega} - 1)}, \tag{9}$$

$$\widetilde{H} t = \frac{\widetilde{\Omega}(\theta - \sin \theta)}{2(\widetilde{\Omega} - 1)^{3/2}}, \tag{10}$$

$$\widetilde{\rho} = \frac{3\widetilde{H}^2 (\widetilde{\Omega} - 1)^3}{\pi G \widetilde{\Omega}^2 (1 - \cos \theta)^3}, \tag{11}$$

with $\widetilde{\Omega}$ and $\widetilde{H}$ being the density parameter and the Hubble constant relative to the overdense shell. The overdense shell reaches maximum expansion for $\theta = \pi$, with the mean density of the shell at maximum expansion being:

$$\widetilde{\rho}_m = 3\pi/32Gt_m^2.$$

Virialization occurs at $t_v = 2t_m$ with a radius $R_v = R_m/2$. The mean density of the overdense sphere at virialization is then:

$$\widetilde{\rho}_v = \frac{3\pi}{Gt_v^2}. \tag{12}$$

Equation (11) can be written as $\widetilde{\rho} = \widetilde{\rho}_v/(1 - \cos\theta)^3$. Since we want to express $\widetilde{\rho}$ as a function of $t_v$, we use equation (12) to obtain:

$$\widetilde{\rho} = \frac{3\pi}{Gt_v^2(1 - \cos\theta)^3}. \tag{13}$$

The parametric equation of the mean density of the universe is:

$$\overline{\rho} = \frac{3H_0^2(1 - \Omega_0)^3}{\pi G \Omega_0^2 (\cosh\psi - 1)^3}. \tag{14}$$

$\Omega_0$ and $H_0$ are the density parameter and the Hubble constant of the universe. The ratio of the perturbation density over the mean background density of the universe is then:

$$\frac{\widetilde{\rho}}{\overline{\rho}} = \frac{\pi^2 \Omega_0^2}{(1 - \Omega_0)^3 (H_0 t_v)^2} \frac{(\cosh\psi - 1)^3}{(1 - \cos\theta)^3}. \tag{15}$$

In this equation we want to express $(\cosh\psi - 1)$ and $(1 - \cos\theta)$ as a function of time. We use the fact that for $t \ll t_0$ (or equivalently, $\psi \ll 1$), the parametric equation of the time,

$$H_0 t = \frac{\Omega_0 (\sinh\psi - \psi)}{2(1 - \Omega_0)^{3/2}}, \tag{16}$$

becomes:

$$H_0 t \sim \frac{\Omega_0}{2(1 - \Omega_0)^{3/2}} \frac{\psi^3}{6} \left(1 + \frac{\psi^2}{20}\right).$$

We invert this equation to obtain $\psi$ as a function of time:

$$\psi^3 \sim \frac{12 H_0 t (1 - \Omega_0)^{3/2}}{\Omega_0} \left(1 - \frac{12^{2/3}(H_0 t)^{2/3}(1 - \Omega_0)}{20\, \Omega_0^{2/3}}\right),$$

to obtain:

$$(\cosh\psi - 1)^3 \sim \frac{12^2 (H_0 t)^2 (1 - \Omega_0)^3}{8\, \Omega_0^2} \times$$

$$\left(1 + \frac{3 \cdot 12^{2/3}(H_0 t)^{2/3}(1 - \Omega_0)}{20\, \Omega_0^{2/3}}\right).$$

Similarly, using the fact that for $t \ll t_0$ ($\psi \ll 1$) equation (10) can be written

$$t \sim \frac{t_v}{2\pi} \frac{\theta^3}{6} \left(1 - \frac{\theta^2}{20}\right),$$

we obtain:

$$(1 - \cos\theta)^3 \sim \frac{12^2 \pi^2 t^2}{8 t_v^2} \left(1 - \frac{3 \cdot 12^{2/3} \pi^{2/3} t^{2/3}}{20 \, t_v^{2/3}}\right).$$

Now, we can use the developments of $(\cosh\psi - 1)^3$ and $(1 - \cos\theta)^3$ as a function of time, to eliminate $\psi$ et $\theta$ from equation (15):

$$\frac{\widetilde{\rho}}{\rho} \sim 1 + \frac{3}{5} \left(\frac{12^{2/3}}{4} \frac{1 - \Omega_0}{\Omega_0^{2/3}} + \frac{12^{2/3}}{4} \frac{\pi^{2/3}}{(H_0 t_v)^{2/3}}\right) (H_0 t)^{2/3}. \tag{17}$$

In this equation, we want to substitute the variable $t$ for the redshift $z$ defined as:

$$\frac{1}{1+z} = \frac{a(t)}{a(t_0)}.$$

From equation (8), and for $\psi \ll 1$:

$$\frac{1}{1+z} \sim \frac{\Omega_0}{2(1-\Omega_0)} \frac{\psi^2}{2}. \tag{18}$$

Using the development of equation (16) until the third order we obtain:

$$(H_0 t)^{2/3} \sim \frac{4}{12^{2/3} \Omega_0^{1/3}} \frac{1}{1+z}.$$

We use this expression of $(H_0 t)^{2/3}$ in equation (17) to finally obtain:

$$\frac{\widetilde{\rho}}{\rho} = 1 + \frac{3}{5} \left(\frac{1 - \Omega_0}{\Omega_0} + \frac{\pi^{2/3}}{\Omega_0^{1/3} (H_0 t_v)^{2/3}}\right) \frac{1}{1+z}.$$

The initial overdensity, $\delta_{c,i}$, of a perturbation which collapses at $t_v$ is:

$$\delta_{c,i}(\Omega_0, t_v) = \frac{3}{5} \left(\frac{1 - \Omega_0}{\Omega_0} + \frac{\pi^{2/3}}{\Omega_0^{1/3} (H_0 t_v)^{2/3}}\right) \frac{1}{1+z_i}.$$

Linearly extrapolated to $z = 0$, this overdensity is

$$\delta_{c,0}(\Omega_0, t_v) = \delta_{c,i} \frac{D(t_0)}{D(t_i)},$$

with $D$ being the linear growth factor given by equation (7). For $t_i \ll t_0$, $D(t_i)$ could be approximated by $D(t_i) \sim \psi_i^2/10$. Using equation (18), $1/((1+z_i)D(t_i))$ becomes: $5\Omega_0/(2(1-\Omega_0))$. Finally:

$$\delta_{c,0}(\Omega_0, t_v) = \frac{3}{2} \frac{\Omega_0}{(1 - \Omega_0)} \left(\frac{1 - \Omega_0}{\Omega_0} + \frac{\pi^{2/3}}{\Omega_0^{1/3} (H_0 t_v)^{2/3}}\right) D(t_0).$$

*2. The density contrast*

The density contrast over the mean density of the background universe of an object which virializes at redshift $z_v$, is the ratio of equation (12) to equation (14):

$$\Delta = \frac{3\pi}{Gt_v^2} \frac{\pi G \Omega_0^2 (\cosh \psi - 1)^3}{3H_0^2 (1-\Omega_0)^3},$$

which can be expressed analytically by mean of equations (8) and (15),

$$\Delta(\Omega_0, z_v) = 4\pi^2 X^3 \times$$
$$\left\{ (X^2 + 2X)^{1/2} - \ln\left(1 + X + (X^2 + 2X)^{1/2}\right) \right\}^{-2},$$

similar to the expression derived by Maoz (1990). In this equation, $X$ is defined as,

$$X = \frac{2(1-\Omega_0)}{\Omega_0} \frac{1}{1+z_v}.$$

In the case $\Omega_0 = 1$, we recover the usual constant value $\Delta = 178$. In the case $\Omega_0 < 1$, $\Delta$ is higher and grows with decreasing the redshift.